
\input harvmac.tex
\vskip 2in
\Title{\vbox{\baselineskip12pt
\hbox to \hsize{\hfill}
\hbox to \hsize{\hfill RU-95-85}}}
{\vbox{\centerline{RR - Dilaton Interaction In a Type IIB Superstring}}}
\centerline{Dimitri Polyakov\footnote{$^\dagger$}
{polyakov@pion.rutgers.edu}}
\medskip
\centerline{\it Dept. of Physics and Astronomy}
\centerline{\it Rutgers University,Piscataway, NJ 08855}
\vskip .5in
\centerline {\bf Abstract}
We analyze the interaction between the massless
 Ramond-Ramond states with a dilaton
in a type II superstring.By  constructing vertex
operators for massless Ramond-Ramond
states and computing their correlation functions with a dilaton  we find
the Ramond-Ramond part of the superstring
low-energy effective action.Those RR terms appear
in the action without the standard dilatonic
factor (string coupling constant) as has
been shown earlier on the basis of  space-time supersymmetry.
The geometrical interpretation of
this fact is presented.Namely we argue that the spin operators
in the RR vertices effectively decrease the Euler character
 of the worldsheet by
1 unit.As a result,the dilaton term in the worldsheet action has the form:
${\sim}\int{d^2}ze^{\Phi}\Lambda{F}{\bar\Lambda}$,
where $\Lambda$ and $\bar\Lambda$ are the $(1,0)$ and
$(0,1)$ spin operators for the ghost and matter
fields,$F$ is the RR field strength
contracted with $\gamma$-matrices.
We also analyse the interaction of a dilaton with the massive RR states.
\Date{October 95}
\vfill\eject
\lref\niehuiz{P.Van Nieuwenhuizen,Phys.Reports {\bf 68-4} (1981) 189}
\lref\shenker{D.Friedan,S.Shenker,E.Martinec,Nucl.Phys.{\bf B271}(1986) 93}
\lref\tzeytlin{M.B.Green,J.H.Schwarz,E.Witten,
Superstring Theory,Cambridge University Press (1987)}
\lref\scherk{J.Scherk,J.H.Schwarz.,Nucl.Physics {\bf B81}(1974)87) Vol.194 N}
\lref\polch{S.Shenker et al,Nucl.Phys. {\bf B316},590 (1989)}
\lref\polchs{J.Polchinski,A.Strominger {\bf UCSBTH-95-30},hep-th 9510227}
\lref\fadd{Faddeev L.D.,Slavnov A. Quantum Theory of Gauge Fields}
\lref\polchino{J.Polchinski,NSF-ITP-95-122,hep-th/9510017}
\lref\ferr{S.Cecotti et.al. Int.Journal of Mod.Physics A V.4,No10(1989),2475}
Recently there has been a significant
progress in the understanding of string-string
dualities.The important step in analyzing the dualities is to study low-energy
effective actions of different string theories.
To obtain the low-energy Lagrangian
for a given string theory,
one has to compute the couplings between its massless states.
In this paper,we analyze how Ramond-Ramond states interact with  a dilaton  in
a closed superstring and derive RR contributions to the effective
Lagrangian directly from superstring theory.
It is of interest to present
string-theoretical derivation of these couplings and to generalise it
to massive
states.
Let us explain the idea of the computation.
The bosonic part of the worldsheet
superstring action perturbed with background fields is given by:
\eqn\lowen{I={1\over{2\pi\alpha}}\int{d^2}z
{1\over2}{\sqrt{\gamma}}{\gamma^{ab}}g_{ij}(X)+
{\epsilon^{ab}}B_{ij}{\partial_a}{X^i}{\partial_b}{X^j}+
{1\over{4\pi}}\int{d^2}z{\sqrt{\gamma}}R^{(2)}\Phi(X)+{V_i}{T_i}(X)}
Here $B_{ij}$ is an axion,$\Phi$ is the space-time
field of a dilaton ,$T_i$ are background space-time fields
and $V_i$ are the corresponding vertex operators.
The corresponding low - energy Lagrangian consists of
the standard NS-NS part:~\refs{\tzeytlin}
\eqn\lowen{L_{NS-NS}{\sim}\int{\sqrt{g}}e^{-2\Phi}
{\lbrack}-R+4\nabla\Phi\nabla\Phi+{1\over{12}}H_{ijk}H^{ijk}{\rbrack},}
where $H_{ijk}=3\partial_{{\lbrack}i}B_{jk{\rbrack}}$,
and the RR part  which is the
sum over the contributions from the antisymmetric field strengths:
\eqn\lowen{L_{RR}={\sum_j}
\int{\sqrt{g}}g^{{\mu_{1}}
{{\mu_{1}}\prime}}...g^{{\mu_{j}}{{\mu_{j}}\prime}}
F_{{\mu_1}...{\mu_j}}F_{{{\mu_1}\prime}...{{\mu_j}\prime}}}
where $F_{{\mu_1}...{\mu_j}}$ are closed $j$-forms and the summation goes
over even $j$ for the type IIA and odd $j$ for the type IIB superstrings.
The well known peculiar feature of the RR sector is the absence of the factor
$e^{-2\Phi}$ in front of the effective
action~\refs{\polchs,\polchino}.It is usually derived from the 10d
supersymmetry ~\refs{\niehuiz}.However, from the string-theoretical point of
view this result is strange since the factor of $e^{-2\Phi}$ is thought to be
universal,coming from the fact that the Euler character of the sphere is
equal to -2.Indeed,under the translation
$\Phi{\rightarrow}{\Phi}+C$(where C is
constant) the expression (1) transforms as $I{\rightarrow}I+C\chi=I-2C$;$\chi$
is the Euler character for our world surface.It follows therefore that the
effective action is multiplied by$e^{C\chi}=e^{-2C}$under this transformation.
 In this paper we shall try to explain  this form of the effective action (3)
from the string-theoretical point of view.In order to obtain effective action
from string theory we have to compare 3-point functions coming from the
computation in string theory with those  from the low-energy Lagrangian.
To do that we have to recast the Lagrangian in the form in which the dilaton
is not  mixing with gravity.Tne necessary field redefinition is well-known
 and is given by:$g^{\mu\nu}{\rightarrow}e^{{-4\Phi}\over{D-2}}g^{\mu\nu}$
after which the action takes the form:
\eqn\grav{\eqalign{L_{NS-NS}{\sim}\int{d^{10}}X{\sqrt{g}}
(-R+4\nabla\Phi\nabla\Phi+{1\over{12}}e^{{8\Phi}\over{D-2}}H_{ijk}H^{ijk})\cr
L_{RR}{\sim}\int{d^{10}}X{\sqrt{g}}(-{1\over4}e^{{3\Phi}\over2}
F_{\mu\nu}F^{\mu\nu}-{1\over{32}}e^{\Phi\over2}S_{\alpha\beta\gamma\delta}
S^{\alpha\beta\gamma\delta}).}}
Here $S_{\alpha\beta\gamma\delta}=
4{\partial_{{\lbrack}\alpha}}S_{\beta\gamma\delta{\rbrack}}$,and
the $3$-form
$C_{\beta\gamma\delta}$ and the vector potential $A_\mu$ constitute
the Ramond-Ramond massless spectrum in a type IIB superstring.
All we have to do now is to extract the ${\Phi}FF$ and ${\Phi}SS$
couplings from string theory.

Note that it is also possible however to recast
this action  into the standard string theory form by redefining the fields.
By using the  redefinition:
$A_{\mu}{\rightarrow}e^{-\Phi}B_{\mu}$,
$C_{\beta\gamma\delta}{\rightarrow}e^{-\Phi}D_{\beta\gamma\delta}$ we may write
the action in the form
\eqn\lowen{{S^{RR}}{\sim}\int{d^10}X{\sqrt{g}}
e^{-2\Phi}(-{1\over4}{{({\partial_{{\lbrack}\mu}}
B_{\nu{\rbrack}}-{\partial_{{\lbrack}\mu}}
\Phi{B_{\nu{\rbrack}}})}^2}-
{1\over{32}}{{({\partial_{{\lbrack}\alpha}}D_{\beta\gamma\delta{\rbrack}}
-{\partial_{{\lbrack}\alpha}}{\Phi}D_{\beta\gamma\delta{\rbrack}})}^2}+...)}
Though the factor
$e^{-2\Phi}$ does appear in the action written in the form (4),the
gauge terms have now a very non-standard form.
Now, the string
coupling $g_{st}$ enters just as in the NS-NS sector.But the price
we pay for that
are the extra cubic and quartic
interactions  between the dilaton and gauge fields which is seen from (5).
Of course the elements of $S$-matrix are invariant under field redifinition
and are the same for (4) and (5).
It is more convenient to compare 3-point functions from string theory
with those following from the Lagrangian (4) containing unrescaled fields.
 So, if we want to check that string theory
really gives (4). we must compute the 3-point functions involving
dilaton and two RR states and show that their value is equal to the one
following from (4).This is what we will do below and then give a geometrical
interpretation of the result.
As we have mentioned  above,the RR sector in the type II superstring theory has
the following massless excitations:
$A_{\mu}$ - the vector;and $C_{\alpha\beta\gamma}$ - the 3-form.
The corresponding vertex operators  must have the form~\refs{\polch}
\eqn\grav{\eqalign{V_{1}(k)=
{1\over2}F_{\mu\nu}(k){{{\lbrack}
{\gamma^\mu},{\gamma^\nu}{\rbrack}}_{AB}}
{\Sigma_A}{{\bar{\Sigma}}_B}e^{-1/2(\phi+{\bar{\phi}})}e^{ikX}\cr
V_{2}(k)={1\over{24}}
S_{\alpha\beta\gamma\delta}(k)
{\gamma^{\lbrack\alpha}}{\gamma^\beta}{\gamma^\gamma}{\gamma^{\delta\rbrack}}
{\Sigma}{{\bar{\Sigma}}}e^{-1/2(\phi+{\bar{\phi}})}e^{ikX}.}}
Here $\Sigma$,$\bar\Sigma$ are the spin operators for matter fields,and $\phi$,
$\bar\phi$ are the bosonized superconformal ghosts.
The requirement of BRST invariance
imposes certain constraints on the polarization tensors in $V_1$ and $V_2$.
To demonstrate this, let us apply the left BRST charge to the first operator
in (6).The BRST invariance condition is
\eqn\lowen{0={\lbrace}Q_{BRST},V_1(k){\rbrace}
=e^{1/2\phi-\chi}\Sigma(\gamma{k})
{\gamma^{\lbrack\mu}}{\gamma^{\nu\rbrack}}F_{\mu\nu}{\bar\Sigma}e^{ikX}}
Here $\chi$ is the auxiliary field
in the bosonization formulas for the superconformal ghosts
$\beta,\gamma$:$\gamma=e^{\phi-\chi}$,$\beta=e^{\chi-\phi}\partial\chi$.
The condition (7) requires therefore that
\eqn\lowen{F_{\mu\nu}(k){\Sigma}(\gamma{k})
{\gamma^{{\lbrack}\mu}}{\gamma^{\nu\rbrack}}\bar\Sigma=0}
This is satisfied if:
\eqn\grav{\eqalign{(k_\alpha{F_{\mu\nu}}+
{k_\mu}F_{\nu\alpha}+{k_\nu}F_{\alpha\mu})=0,\cr{k_\mu{F_{\mu\nu}}=0}}}
-the  Maxwell's equations.
{}From these equations it follows that $F_{\mu\nu}$ can be represented as
 $F_{\mu\nu}(k)=1/2(e_\mu(k)k_\nu-e_\nu(k)k_\mu)$where the polarization
vector satisfies $(e(k)k)=0$.
Analogously,the BRST condition
for $V_2$  fixes $S_{\alpha\beta\gamma\delta}(k)
={1\over{24}}k_{\lbrack\alpha}s_{\beta\gamma\delta\rbrack}(k)$
where $s_{\beta\gamma\delta}$ is some antisymmetric tensor satisfying
the transversality condition:$k_\beta{s_{\beta\gamma\delta}}=0$.
Therefore the BRST-invariant type IIB RR massless vertices are
\eqn\grav{\eqalign{e_\mu(k)V^\mu(k)=e_\mu(k)
e^{-1/2(\phi+\bar\phi)}\Sigma(\gamma{k}){\gamma^\mu}{\bar\Sigma}e^{ikX}\cr
s_{\alpha\beta\gamma}(k)V^{\alpha\beta\gamma}
=1/6s_{\alpha\beta\gamma}(k)(\gamma{k})
{\gamma^{\lbrack\alpha}}{\gamma^\beta}{\gamma^{\gamma\rbrack}}e^{ikX}}}
{}.
The equivalent formulas for the massless RR vertices have been proposed
in  ~\refs{\polch}(see also~\refs{\polchino})
Now that we have the vertices corresponding to the RR states,
let us compute their  couplings with the  dilaton vertex $V_\Phi$.
The vertex operator of the dilaton is given by~\refs{\scherk}
${V_{\Phi}^{(0)}}=e_{\mu\nu}(k){\partial}{X^\mu}{\bar\partial}{X^\nu}e^{ikX}$,
where ${e_{\mu\nu}}(k)={\eta_{\mu\nu}}-k_\mu{\bar{k_\nu}}-{\bar{k_\mu}}k_\nu$.
Here $\eta_{\mu\nu}$ is the flat space-time metric and ${\bar{k}}$ is defined
so that ${{\bar{k}}^2}=0$,$(k{\bar{k}})=1$.
For our computations it will be more convenient to use
the dilaton emission vertex in another picture: ${V_\Phi}={e_{\mu\nu}}(k)
e^{-\phi-\bar\phi}{\psi^\mu}{{\bar\psi}^\nu}$.
The $V_{\Phi}^{(0)}$
is obtained by the left and right picture-changings of $V_\Phi$.
The correlation function of 2 massless vector gauge particles with a dilaton is
\eqn\grav{\eqalign{{e_\mu}(k_1){e_\nu}(k_2)
({\eta_{\alpha\beta}}-k_\alpha{\bar{k_\beta}}-{\bar{k_\alpha}}k_\beta)
<V_\mu(k_1)V_\nu(k_2)e^{-\phi-{\bar\phi}}{\psi^\alpha}{\bar{\psi^\beta}}>=\cr
=<e^{-1/2(\phi+\bar\phi)}{\Sigma_A}{\bar{\Sigma_B}}{\gamma^\mu}
({\gamma}k_1)e^{ik_1{X}}(z_1,{\bar{z_1}})e^{-1/2(\phi+\bar\phi)}{\Sigma_C}{\b
ar{\Sigma_D}}{\gamma^\nu}(\gamma{k_2})e^{ik_2{X}}(z_2,{\bar{z_2}}){\times}\cr
{\times}e^{-\phi-\bar\phi}{\psi_\alpha}
{\bar{\psi_\beta}}e^{ikX}(z_3,{\bar{z_3}})>{e^\mu}(k_1){e^\nu}(k_2)
({\eta_{\alpha\beta}}-k_\alpha{{\bar{k}}_\beta}-{{\bar{k}}_\alpha}k_\beta)=\cr
={1\over{(z_1-z_2)(z_2-z_3)(z_3-z_1)({\bar{z_1}}-{\bar{z_2}})({\bar{z_2}}-
{\bar{z_3}})({\bar{z_3}}-{\bar{z_1}})}}e_\mu(k_1)e_\nu(k_2){\times}\cr
{\times}({\eta_{\alpha\beta}}-k_\alpha{\bar{k_\beta}}-
{\bar{k_\alpha}}{k_\beta})Tr{\lbrack}({\gamma}k_1){\gamma^\mu}
{\gamma^\alpha}({\gamma}k_2){\gamma^\nu}{\gamma^\beta}{\rbrack},}}
with $k=-k_1-k_2$.
The contribution to the 3-point function is therefore
\eqn\grav{\eqalign{{C_{{A_\mu}
{A_\nu}\Phi}}=1/2Tr{\lbrack}({\gamma}k_1)({\gamma}e(k_1))
{\gamma^\alpha}({\gamma}k_2)({\gamma}e(k_2)){\gamma_\alpha}{\rbrack}-\cr
-1/2Tr{\lbrack}({\gamma}k_1)({\gamma}e(k_1))
({\gamma}k)({\gamma}k_2)({\gamma}e(k_2))({\gamma}{\bar{k}}){\rbrack}-\cr
-1/2Tr{\lbrack}({\gamma}k_1)({\gamma}e(k_1))
({\gamma}{\bar{k}})({\gamma}k_2)({\gamma}e(k_2))({\gamma}k){\rbrack}}}
The first ,second and third terms in (7) contribute
\eqn\grav{\eqalign{{C_{{A_\mu}
{A_\nu}\Phi}^{(1)}}=1/2Tr{\lbrack}
(\gamma{k_1})(\gamma{e(k_1)}){\gamma^\alpha}
(\gamma{k_2})(\gamma{e(k_2)}){\gamma_\alpha}{\rbrack}=\cr=
-1/2Tr{\lbrack}(\gamma{e(k_1)})
(\gamma{k_2})(\gamma{e(k_2)})(\gamma{k_1}){\rbrack}+
1/2Tr{\lbrack}(\gamma{e(k_1)})(\gamma{k_1})(\gamma{k_2})
{(\gamma{e(k_2)})}{\rbrack}-\cr-1/2({k_1}e(k_2))
Tr{\lbrack}(\gamma{e(k_1)}){\gamma^\alpha}(\gamma{k_2})
{\gamma_\alpha}{\rbrack}=\cr
=-1/2((e(k_1)k_2)(e(k_2)k_1)-(e(k_1)k_2)(e(k_2)k_1)-\cr-(e(k_2)k_1){\lbrack}
2e(k_1)k_2-D(e(k_1)k_2){\rbrack})=1/2(D-4)({k_1}e(k_2))({k_2}e(k_1))\cr
{C_{{A_\mu}{A_\nu}\Phi}^{(2)}}={C_{{A_\mu}{A_\nu}\Phi}^{(3)}}=
-1/2Tr{\lbrack}(\gamma{k_1})(\gamma{e(k_1)})(\gamma{k})(\gamma{k_2})
(\gamma{e(k_2)})(\gamma{\bar{k}}){\rbrack}=\cr=1/2(k_1{
e(k_2)})Tr{\lbrack}(\gamma{e(k_1)})(\gamma{k})(\gamma{k_2})
(\gamma{\bar{k}}){\rbrack}-1/2(k_1{\bar{k}})Tr{\lbrack}(\gamma
{e(k_1)})(\gamma{k})(\gamma{k_2})(\gamma{e(k_2)}){\rbrack}=\cr=
1/2(k_1{e(k_2)}){\lbrack}-(k_2{e(k_1)})
(k_2{\bar{k}})-(k_2{e(k_1)})
(k{\bar{k}}){\rbrack}-\cr-1/2(k_1{\bar{k}})(e(k_1)k_2)(e(k_2)k_1)=0}}
therefore
\eqn\lowen{{C_{{A_\mu}{A_\nu}\Phi}}=3({k_1}e(k_2))({k_2}e(k_1))}
Note that for $D=4$ this 3-point function would be zero, as has been shown
in ~\refs{\ferr}
Hence the corresponding term in
the effective Lagrangian will be ${\sim}3/8{F_{\mu\nu}}{F^{\mu\nu}}\Phi(X)$.
Another 3-point function contributing to the interaction of a dilaton
with the massless RR states is that of
$V_\Phi$ with 2 gauge 3-forms.By using the vertex operators (10) we write
the correlation function as
\eqn\grav{\eqalign{s_{\alpha\beta\gamma}(k_1)s_{ijk}
(k_2)<V^{\alpha\beta\gamma}(z_1,k_1)V^{ijk}(z_2,k_2)V_{\Phi}(z_3,k_3)>=\cr
={1\over{36}}s_{\alpha\beta\gamma}(k_1)
s_{ijk}(k_2){{[({\gamma}k_1){\gamma^{[\alpha}}{\gamma^\beta}
{\gamma^{\gamma]}}]}_{AB}}{{[({\gamma}k_2)
{\gamma^{[i}}{\gamma^j}{\gamma^{k]}}]}_{CD}}
({\eta_{\mu\nu}}-{k_\mu}{\bar{k_\nu}}-{\bar{k_\mu}}{k_\nu}){\times}\cr
{\times}<e^{-1/2(\phi+\bar\phi)}
{\Sigma_A}{{\bar{\Sigma}}_B}e^{i{k_1}X}
(z_1,{\bar{z_1}})e^{-1/2(\phi+\bar\phi)}
{\Sigma_C}{{\bar{\Sigma}}_D}e^{i{k_2}X}(z_2,{\bar{z_2}}){\times}\cr
{\times}e^{-\phi-\bar\phi}{\psi^\mu}{{\bar{\psi}}_\nu}
e^{ikX}(z_3,{\bar{z_3}})>}}
The 3-point function is hence given by
\eqn\grav{\eqalign{C_{cc\Phi}={1\over{36}}
{e_{\mu\nu}}(k)s_{\alpha\beta\gamma}(k_1)
s_{ijk}(k_2)1/2Tr{\lbrack}({\gamma}k_1){\gamma^{{\lbrack}\alpha}}
{\gamma^\beta}{\gamma^{\gamma{\rbrack}}}{\gamma^\mu}({\gamma}k_2)
{\gamma^{{\rbrack}i}}{\gamma^j}{\gamma^{k{\rbrack}}}{\gamma^\nu}{\rbrack}=\cr
={1\over{72}}s_{\alpha\beta\gamma}(k_1)s_{ijk}(k_2)Tr{\lbrack}
({\gamma}k_1){\gamma^{{\lbrack}\alpha}}
{\gamma^\beta}{\gamma^{\gamma{\rbrack}}}{\gamma^\mu}
({\gamma}k_2){\gamma^{{\lbrack}i}}{\gamma^j}{\gamma^{k{\rbrack}}}
{\gamma_\mu}{\rbrack}-\cr-{1\over{72}}s_{\alpha\beta\gamma}
(k_1)s_{ijk}(k_2)Tr{\lbrack}({\gamma}k_1){\gamma^{{\rbrack}\alpha}}
{\gamma^\beta}{\gamma^{\gamma{\rbrack}}}({\gamma}k)({\gamma}k_2)
{\gamma^{{\lbrack}i}}{\gamma^j}{\gamma^{k{\rbrack}}}({\gamma}{\bar{k}})
{\rbrack}-\cr-{1\over{72}}s_{\alpha\beta\gamma}(k_1)s_{ijk}(k_2)
Tr{\lbrack}({\gamma}k_1){\gamma^{{\lbrack}\alpha}}{\gamma^\beta}
{\gamma^{\gamma{\rbrack}}}({\gamma}{\bar{k}})
({\gamma}k_2){\gamma^{{\lbrack}i}}{\gamma^k}
{\gamma^{l{\rbrack}}}({\gamma}k){\rbrack}=\cr=
{1\over{72}}({{C_{cc\Phi}}^{(1)}}+{{C_{cc\Phi}}^{(2)}}+{{C_{cc\Phi}}^{(3)}})}}
The 3 terms  in the r.h.s. contribute:
\eqn\grav{\eqalign{{C_{cc\Phi}^{(1)}}= 72{k_{1\alpha}}
{{s_{\delta\beta\gamma}}}(k_1){k_{2\delta}}{{s_{\alpha\beta\gamma}}}(k_2);\cr
{C_{cc\Phi}^{(2)}}={C_{cc\Phi}^{(3)}}=0;}}
The  antisymmetrization over $\alpha,\beta,\gamma,\delta$ is implied.

The 3-point function
 (15)  is equal to
\eqn\lowen{{C_{cc\Phi}}=(k_{1{\lbrack}\alpha}s_{\delta\beta\gamma{\rbrack}}
(k_2))(k_{2{\lbrack}\delta}s_{\alpha\beta\gamma{\rbrack}}(k_2))}
The corresponding contribution to the low-energy effective
space-time  Lagrangian then will  be ${\sim}{1\over{64}}
{S_{\alpha\beta\gamma\delta}}{S^{\alpha\beta\gamma\delta}}\Phi(X)$,
where $S_{\alpha\beta\gamma\delta}(X)=4{\partial_{{\lbrack}\alpha}}
C_{\beta\gamma\delta{\rbrack}}(X)$.

Now that we know from (14),(18) the couplings
of the massless RR states with a dilaton it
is easy to write down the massless RR-terms interacting
with a space-time dilaton in the low-energy effective action:
\eqn\grav{\eqalign{{S_{eff}^{RR\Phi}}\sim{\int}{d^{10}}X
{\sqrt{g}}(-{1\over4}e^{3/2\Phi}F_{\mu\nu}F^{\mu\nu}-{1\over{2}}
e^{1/2\Phi}{{({\partial_{{\lbrack}\alpha}}
{C_{\beta\gamma\delta{\rbrack}}})}^2})
=\cr=\int{d^{10}}X{\sqrt{g}}({-1\over4}e^{3/2\Phi}
F_{\mu\nu}F^{\mu\nu}-{1\over{32}}e^{1/2\Phi}
S_{\alpha\beta\gamma\delta}S^{\alpha\beta\gamma\delta})}}
We see therefore that the string
perturbation theory gives the same effective
Lagrangian as in (4),that followed from the 10d supersymmetry arguments.
By making the inverse rescaling we may
return to the effective action written in the form (3).Thus, we have shown
that the string perturbation theory does give the low-energy limit
with the factor $e^{-2\Phi}$ absent in the Ramond-Ramond effective action.

Let us try to give the geometrical interpretation to these results.
We will argue that the correct contribution of the RR states to the
worldsheet action is given by:
\eqn\lowen{{I_{RR}}{\sim}\int{d^2}z{e^{\Phi(X)}}\Lambda
({\gamma^{{\lbrack}\mu}}{\gamma^{\nu{\rbrack}}}F_{\mu\nu} + ...){\bar\Lambda}}
where $\Lambda$ is a total
(ghost $+$ matter) spin operator~\refs{\shenker}:$\Lambda={\Sigma_{gh}}\Sigma$
 and F is RR field
strength.This would give an extra factor $e^{2\Phi}$ in the target space action
and the formula (3).The origin of  the $e^{\Phi}$ factor can be interpreted as
follows.When we define the RR state we have to cut a
small hole in the surface and to require fermions
to be antiperiodic around this hole.
Cutting the hole reduces Euler character by 1 unit.
Hence we conjecture that any insertion of the
RR state must be accompanied by the factor $e^\Phi$ in the worldsheet action.
The absence of $g_{st}$ in (3) is now simply
explained by the fact that a sphere with 2 holes has zero Euler character.
The formula (20) has two important consequences.First of all,
we should expect that  higher order corrections
to the RR effective action     should
multiplied by the various  dilatonic factors
${\sim}e^{(n-2)\Phi}$ where n is the number of insertions of the spin operators
$\Sigma$(and $\bar\Sigma$)
in the correlation functions contributing to the corresponding higher
order corrections to the $\beta$-function.
Next,the terms corresponding to the coupling of massive RR states with
the dilaton $\Phi$ should  also (in the lowest order) appear in the action
without any dilatonic factor, exactly as in the case of the massless
RR states considered above.Note that it is still possible to apply the string
perturbation theory to
the massive RR - dilaton
interactions since the field of the dilaton is changing slowly.
Let us check the proposal regarding the massive RR- dilaton couplings in
the case of the massive RR states at the lowest level with $k^2=2$ (the lowest
massive level).
Since in this case we have $dim(e^{ikX})=-1$,
the most general gauge-invariant expression
for the RR operator of dimension 1 with $k^2=2$ is
\eqn\grav{\eqalign{V^{{\mu_1}...{\mu_{2n-1}}\alpha\beta}
={1\over{(2n-1)!}}e^{-1/2(\phi+\bar\phi)}
{\Sigma}{\gamma^{{\lbrack}{\mu_1}}}...
({\gamma^{\mu_{2n-1}}}{\gamma^{{\mu_{2n}}{\rbrack}}}
{k_{\mu_{2n}}}-{k^{\mu_{2n-1}}{\rbrack}}){\bar\Sigma}{\times}\cr
{\times}({\partial}{X^\alpha}+iA(k\psi)\psi^\alpha)
(\bar\partial{X^\beta}-iB(k\bar\psi){\bar\psi}^{\beta})e^{ikX}}}
where $A$ and $B$ are some numbers.The condition of the BRST invariance
fixes $A=B=1$.For simplicity,let us consider the case of the emission of a
massive (with $k^2=2$) vector particle $D_\mu(X)$ and compute the interaction
of 2 gauge particles with the dilaton $\Phi$.
It follows from (21) that
the vertex
operator of the emission of a massive $(k^2=2)$ vector boson is given by
\eqn\grav{\eqalign{{V_D^\mu(k)}=
e^{-1/2(\phi+\bar\phi)}{\lbrack}{1\over8}
{\Sigma_A}{{({\gamma^\mu}(\gamma{k})-k^\mu})_{AB}}{{\bar\Sigma}_B}
(\partial{X^\alpha}+i(k\psi)\psi^\alpha)(\bar\partial{X_\alpha}-i
(k\bar\psi)\bar\psi_\alpha)e^{ikX}+\cr+\Sigma((\gamma{k}){\gamma^\alpha}
-k^{\alpha}){\bar\Sigma}(\partial{X_\alpha}+i(k\psi)\psi_\alpha)
(\bar\partial{X^\mu}-i(k\bar\psi){{\bar\psi}^\mu})e^{ikX}+\cr+\Sigma
((\gamma{k}){\gamma^\beta}-k^{\beta}){\bar\Sigma}
(\partial{X^\mu}+i(k\psi){\psi^\mu})
(\bar\partial{X_\beta}-i(k\bar\psi){\bar\psi_\beta})e^{ikX}{\rbrack}}}
The  computation of the relevant 3-point
correlation function gives the following contribution to the $\beta$-function:
\eqn\grav{\eqalign{{1\over2}(z_1-z_2)(z_2-z_3)
(z_3-z_1)(\bar{z_1}-\bar{z_2})({\bar{z_2}}-\bar{z_3})
({\bar{z_3}}-{\bar{z_1}}){\times}\cr{\times}e_\mu(k_1)e_\nu(k_2)<{V_D^\mu}
(k_1){V_D^\nu}(k_2)V_{\Phi}>=
{3\over2}(k_1{e(k_2)})(k_2{e_(k_1)})+2(e(k_1)e(k_2)).}}
The contribution to the effective Lagrangian is therefore
$\sim({3\over2}({k_1}D(k_2))({k_2}D(k_1))+2D^2)\Phi$.
We see that the coupling (23) is reproduced by the following terms in the
effective action:
\eqn\grav{\eqalign{{S^{DD\Phi}_{eff}}
{\sim}-\int{d^{10}}X {\sqrt{g}}
({1\over4}e^{3/2\Phi}{D_{\mu\nu}}
D^{\mu\nu}(X)+e^{2\Phi}{D^2}(X))=\cr=-
\int{d^{10}}X{\sqrt{g}}({1\over4}e^{3/2\Phi}
D_{\mu\nu}D^{\mu\nu}(X)+e^{2\Phi}{{\mu}\over2}{D^2}(X))}}
Here $D_{\mu\nu}$ is the
massive gauge field strength;$\mu=k^2$ is the mass of the gauge vector boson.
Again, after the inverse rescaling of the metric tensor
$g_{\mu\nu}{\rightarrow}e^{-{{4\Phi}\over{D-2}}}g_{\mu\nu}$
in order to bring the kinetic term
to the standard form as in (2)  we find that
\eqn\lowen{{S^{DD\Phi}_{eff}}{\sim}-\int{d^{10}}
X{\sqrt{g}}({1\over{4}}D_{\mu\nu}D^{\mu\nu}(X)+{{\mu}\over2}{D^2}(X)),}
i.e. we see that the massive vector boson  with $k^2=2$ is coupled with a
dilaton in a non-standard way as in (3),as we expected.
The straightforward proof of this proposal for other massive RR modes (tensors
of higher ranks and with bigger masses) is principally
the same though it requires much more
cumbersome calculations.We hope to perform these computations in future papers,
as well as the computations for the higher loops in the RR sector.
\centerline{\bf Conclusion}
We have shown that,in the lowest order of the string perturbation theory
 both massless and some massive Ramond-Ramond states have unusual coupling
with a dilaton..This is related to the factor $e^{\Phi(X)}$
in the worldsheet action of the dilaton that has been found in (20).Just like
the term ${\sim}R^{(2)}\Phi(X)$ in  (1) appears as after the
normal ordering of the vertex operator of a graviton,the worldsheet term (20)
should also be "generated" by the normal ordering of a certain generally
covariant vertex operator in the superstring theory which is yet unknown.
To find the expression for this vertex operator would be very helpful in
our understanding of the string perturbation theory in case
of the supersymmetric background.Also, the conjecture about the relation of
the dilatonic factors
to the insertions of Ramond-Ramond states should be proved in detailes.
 It should therefore be necessary  to consider higher order corrections
in the Ramond-Ramond sector.
\centerline{\bf Acknowledgements}
I'm very thankful to D.Friedan,A.M.Polyakov and S.Shenker for
many helpful comments.The work was supported
from the Grant of the High Energy Theory group of Rutgers University.
\listrefs

\vskip .5in
\footatend\vfill\immediate\closeout\rfile\writestoppt
\centerline{\bf{References}}\bigskip{\frenchspacing%
\parindent=20pt\escapechar=` \input
\jobname.refs\vfill\eject}\nonfrenchspacing

\end